# Mechano-lithography: stress anisotropy driven nematic order in growing three-dimensional bacterial biofilms

**Authors:** Changhao Li[1], Japinder Nijjer[4], Luyi Feng[1], Qiuting Zhang[4], Jing Yan[4,5,*], Sulin Zhang[1,2,3*]

**Affiliations:**

[1]Department of Engineering Science and Mechanics, Pennsylvania State University, University Park, PA, USA.

[2]Department of Biomedical Engineering, Pennsylvania State University, University Park, PA, USA.

[3]Department of Material Science and Engineering, Pennsylvania State University, University Park, PA, USA.

[4]Department of Molecular Cellular and Developmental Biology, Yale University, New Haven, CT

[5]Quantitative Biology Institute, Yale University, New Haven, CT

*Corresponding author. Email: jing.yan@yale.edu or suz10@psu.edu

**Abstract:** Living active collectives have evolved with remarkable self-patterning ability to meet the physical and biological constraints for growth and survival. However, how complex multicellular patterns emerge from a single founder cell remains elusive. Here, by recourse to an agent-based model, we track the three-dimensional (3D) morphodynamics and cell orientational order of growing bacterial biofilms encased by agarose gels. Confined growth causes spatiotemporally heterogeneous stress buildup in the biofilm. High hydrostatic and low shear stresses at the core of the biofilm promote viscous-to-elastic transition and randomize cell packing, whereas the opposite stress state near the gel-cell interface drives nematic ordering with a time delay inherent to shear stress relaxation. Overall, stress anisotropy spatiotemporally coincides with nematic order in the confined biofilms, suggesting an anisotropic-stress-driven ordering mechanism. The strong reciprocity between stress anisotropy and cell ordering inspires innovative 3D mechano-lithography of living active collectives for a variety of environmental and biomedical applications.

**One-Sentence Summary**: Bacterial biofilms is spatiotemporally patterned by growth-induced anisotropic mechanical stresses.

**Main Text:**

**Introduction**

Biofilms are surface-attached aggregates of microorganisms in which bacterial cells are embedded in a complex three-dimensional (3D) polymeric matrix (*1*). In harsh microenvironments, bacteria actively seek survival niches and develop into densely packed biofilms. The constituent rod-shaped bacterial cells within the biofilms often exhibit long-range nematic order (*2-4*), resembling other ordered living active collectives at different scales (*5-7*). While high-density, aligned packing could potentially afford bacterial cells protection from the detrimental chemicals in host and natural environments (*8*), it may also critically underlie the percolation pathways for delivery of nutrients, oxygens, and other molecules essential to biofilm development and fitness (*9, 10*). However, how the long-range order emerges from a single founder cell in its course to a mature 3D biofilm remains poorly understood. A fundamental understanding of cell alignment may provide insights into the biomechanics of biofilm development and shed light on 3D self-patterning of living active-matter systems.

Various techniques have been used to control biofilm morphodynamics, including optically controlled gene expression (*11*), patterned substrates (*12*), and microfluidic devices (*13*). In all these cases, mechanical forces, either passively applied or actively generated, have emerged as a key factor regulating the morphological evolution and internal cell ordering of bacterial communities (*14, 15*). Indeed, it has been shown that hydrostatic, adhesive, and fluid shear forces could all contribute to the 3D morphology and microscopic ordering of biofilms (*16-20*). Here, we develop an agent-based model to recapitulate the growth dynamics of biofilms and uncover the ordering mechanisms underlying the developmental patterns. We reveal a strong spatiotemporal reciprocity between growth-induced stress anisotropy and orientational cell ordering. On one hand, cell proliferation in a confined visco-elastic medium causes a continuous buildup of spatiotemporally heterogeneous and anisotropic mechanical stresses in the growing biofilm. On the other, the growth-induced stress morphs the biofilms, generating anisotropic morphology and long-range nematic order in the biofilms. Such a reciprocity between stress anisotropy and cell ordering inspires mechano-lithographic strategies for biofilm organizations that are applicable to other living active collectives.

**Results**

**Confinement stiffness regulates morphology and cell order in growing biofilms**

We have previously established single-cell microscopy to image the growth dynamics of *V. cholerae* biofilms from a single founder cell to a mature 3D community at single-cell resolution (*20, 21*). We recapitulate the experimental setup and observation below: The growing biofilms were either fully embedded within an agarose gel or confined between a glass surface and an agarose gel. We hereafter term the former geometry-I (G-I) and the latter geometry-II (G-II) biofilms, respectively. We found that the morphodynamical evolution of the biofilms strongly depends on the degree of mechanical confinement (**Fig. 1, A-B**), controlled by the agarose gel stiffness (Young's modulus: 0.1-100 kPa). Over time, the biofilms embedded in soft gels grew into a spherical shape, whereas the biofilms encased by stiff gels were nearly isotropic at the initial stage of development, but gradually attained morphological anisotropy as growth continued. Meanwhile, with increasing gel stiffness, mature G-I biofilms underwent a spherical-to-ellipsoidal shape transition (**Fig. 1A, C-D**), while the G-II biofilms a dome-to-lens shape transition (**Fig. 1B, E-F**). Taken together, confined biofilm growth and gel stiffness jointly regulate the morphological anisotropy of mature biofilms.

The overall shape transitions are concomitant with heterogeneous cell packing and orientational cell ordering inside the biofilms. To quantify cell alignment, we measured the nematic order parameter $S$, defined as the largest eigenvalue of the traceless tensor $\boldsymbol{Q} = \frac{1}{2} \langle 3\hat{\boldsymbol{n}}_{c,i} \otimes \hat{\boldsymbol{n}}_{c,i} - \boldsymbol{I} \rangle$, where $\hat{\boldsymbol{n}}_{c,i}$ is the director of the cell $i$, $\boldsymbol{I}$ is the identity tensor and the angled bracket denotes averaging in a subset of the biofilm (*22*). $S$ ranges from 0 for a completely random state to 1 for a fully nematically ordered state. For biofilms with isotropic morphologies, the constituent cells exhibited a disordered phase. For biofilms with anisotropic morphologies, a nematically ordered, bipolarly aligned layer emerged at the cell-gel interface, reminiscent of liquid crystal droplets (*23, 24*). Such 3D cell alignment became more pronounced with increasing gel stiffness. As cells continued to grow and divide, the thickness of the nematically ordered phase increased **(Fig. 1, C-F)**, which was successfully captured by our simulations **(Fig. 1, G-J)**. The similar nematic order of the differentially confined biofilms suggests a robust, self-patterning mechanism of the growing biofilms, which is quantitatively dissected below.

**Agent-based model recapitulates the 3D morphology and cell ordering**

To better understand the driving force for orientational cell ordering, we developed an agent-based model (ABM), where the gel is modeled as spherical agents and individual bacterial cells as a growing/dividing spherocylinder with a hard core and a soft shell. Cell-cell and cell-gel interactions are described by a modified Johnson-Kendall-Roberts (JKR) model (*25, 26*) and the gel-gel interactions by a harmonic potential. The inter-agent interactions capture the material properties of the biofilm, the gel, and the interfaces (Supplementary Materials). The interaction parameters are calibrated by experimental measurements, which can be tuned to match different bacterial mutants with varying cell-cell, cell-substrate, and cell-gel interactions. With the tailored cell-gel interactions, our ABM can treat environmental confinement in a unified platform and extends the application of the model. The hard-core, soft-shell bacterial cell agents not only capture the extracellular-matrix-mediated elasticity in biofilms, but also avoid unphysical penetration of cells into the gels under high pressure, in contrast to other ABMs (*17, 27, 28*) where bacterial cells are modeled by rigid rods. ABMs can also quantitively track the forces acting on all the constituent agents, which are inaccessible to experimental imaging.

Our ABM accurately recapitulates the growth dynamics of biofilms from early time to mature biofilm communities (**Fig. 1**, **Fig. S2**). For the G-II biofilms, starting with a single bacterial cell lying on the glass surface, continuous cell division and growth first form a monolayer on the basal plane. The local in-plane pressure accumulates due to the dense packing of cells. When the pressure reaches a threshold, cells near the center undergo mechanical instability whereby they rotate to a vertical orientation, partially releasing the in-plane growth pressure (*17*). This verticalization process initializes out-of-plane growth and the biofilm transits from 2D to 3D colonies, which agrees with multiple previous experimental and theoretical works (*2, 29-32*). In contrast, the G-I biofilms are nearly isotropic in the initial stage of development but become anisotropic as cells continue to grow and divide. Thus, our ABM captures the gel stiffness-dependent shape transitions as well as cell alignment for both the G-I and G-II biofilms (**Fig. 1, G-J, Fig. S2**). The ABM thus offers a unique platform to dissect morphological change and cell ordering mechanisms in mechanically confined biofilms.

**Confined growth imposes spatial stress heterogeneity and anisotropy in biofilms**

Cell proliferation adds volume into the biofilm, generating growth-induced stress under the confinement. To quantify the spatiotemporal stress distribution inside the growing biofilms, we calculated the Virial stress (*33*) ($\boldsymbol{\sigma}$) by the contribution of particle interactions and the viscosity from surrounding environments (Supplementary Materials). Our simulations revealed that the biofilm growth causes an exponential buildup of the average hydrostatic pressure $p = \frac{1}{3}\text{tr}(\boldsymbol{\sigma})$ (**Fig. 2A**), due to the exponentially increasing cell number *N* and volume. In contrast, the average equivalent shear stress, $\tau_{eq} = \sqrt{2\boldsymbol{\tau}:\boldsymbol{\tau}/3}$ (where $\boldsymbol{\tau} = \boldsymbol{\sigma} - p\mathbf{I}$ is the deviatoric stress), increases initially but plateaus at a later stage, indicating a time-dependent shear relaxation process. The average pressure and shear stress increases monotonically with gel stiffness (**Fig. 2B**), suggesting that gel confinement contributes to the stress buildup. In space, the hydrostatic stress gradually decreases from the biofilm core to the cell-gel interface, suggesting a decreasing degree of local volumetric confinement (**Fig. 2, C-D**). Oppositely, shear stress increases from the core to the interface (**Fig. 2, E-F**), which further evidences that shear stress arises from the cell-gel adhesion. The shear stress pattern resembles the boundary layer in viscous flows, where shear stress linearly decays from the boundary (*34*) (**Fig. S4**). Such spatial stress distributions emerge over time as biofilms grow (**Fig. 2, D and F**), concomitant with cell orientational ordering (**Fig. 1, F and J**).

**Spatiotemporally varying stress facilitates heterogeneous cell reorientation**

In stress-free conditions, bacterial cell aggregates are viscous alike (flows like a fluid) due to their relatively large cell-cell distances (*2, 6, 24*), which allows individual bacteria to swim and rotate under viscous force. In contrast, when bacteria transition to form biofilms, they display elastic behavior (deform like a solid), due to the production of extracellular matrices and dense cellular packing (*35, 36*). Under mechanical confinement, the exponential buildup of hydrostatic pressure inside a growing biofilm further densifies cell packing (**Fig. 3A**), promoting a viscous-to-elastic transition (VET). Kymograph of the biofilm density reveals the progressive densification of the core with biofilm growth (**Fig. 3B**). Consistent with the spatial variation of the hydrostatic pressure, biofilm density decays from the core to the gel-cell interface (**Fig. 3C**). Densified biofilms correspond to higher resistance to the rotation of rod-shaped cells, potentially suppressing cell alignment in biofilms.

Nematic order requires cell rotation and reorientation, which in turn, needs to overcome viscous and elastic stresses in the biofilms. Shear stress causes cell rotation, whereas shear resistance scales with the hydrostatic pressure, according to the Mohr-Coulomb criterion (*37*) (namely, higher pressure increases the energy barrier for shear-induced cell rotation). Indeed, we found that cell rotation speed increases with the equivalent shear stress normalized by the hydrostatic pressure $\tau_{eq}/p$ (**Fig. 3D**). At the early stage of the biofilm growth (~6 hrs.), the biofilms behave more like viscous fluids due to the low hydrostatic stress; however, the cells still have a relatively small rotation speed because of low shear stress, and thus a generally disordered phase. As the biofilm continues to grow, the hydrostatic stress at the core of the biofilm increases (**Fig. 2, C-D**), leading to the VET of the biofilm core. Near the cell-gel interface, the hydrostatic stress is relatively low and the shear stress transmitted from the biofilm-gel interface (*20*) is relatively high. Thus, the biofilm near the cell-gel interface remained its viscous characteristic, manifested by the relatively low density and high rotation speed (**Fig. 3, E-F**). This leads to increased cell alignment at the biofilm-gel interface. The increasing shear resistance and decreasing shear stress from the interface to the core promote simultaneous VET and nematic order-to-disorder transition.

**Stress anisotropy drives nematic ordering**

The G-I biofilm grows with an increasing nematic order ($S$) of the bacterial cells (**Fig. 4A**). The bipolar order ($S_b$), characterizing the degree of cell alignment with its local meridian (supplementary materials) also increases. To elucidate how the growth-induced stress drives bacterial cell ordering, we note that a bacterial cell may experience a general triaxial compressive stress state described by the three principal stresses, $0 \geq \sigma_1 \geq \sigma_2 \geq \sigma_3$ (**Fig. S9**). For a bacterial cell with a rod shape, under hydrostatic compression, elastic strain energy arises predominately from the compression along its long axis (*38*). Thus, the most energetically stable state is when the director of the cell $\hat{n}_c$ aligns with the direction of the minimal compressive stress, i.e., the *direction of the first principal stress*, $\hat{n}_1$. For a bacterial cell that orients differently from $\hat{n}_1$, a thermodynamic driving force exists to drive the cell to reorient toward the $\hat{n}_1$ direction to lower the local elastic energy. To further confirm that cells have the tendency to reorient toward $\hat{n}_1$, we defined a relative orientation parameter $\alpha = |\hat{n}_1 \cdot \hat{n}_c|$. Here $\alpha = 1$ describes the state at which the cell fully aligns with the first principal stress. We found that as the biofilm grows, the spatially averaged value of $\alpha$ monotonically increases (**Fig. 4B**), suggesting the growth stress patterns the cells (**Fig. 4C**). In space, the alignment parameter $\alpha$ also increases monotonically from the core to the gel-cell interface (**Fig. 4, B-C, Fig. S6**), suggesting most alignments occur near the gel-cell interface where the shear stress is largest. The first principal stress near the interface is radially directed along the interface, stabilizing a bipolar alignment of the cells (**Fig. 4D**).

To investigate why cells at the core of the biofilms do not follow the $\hat{n}_c$ direction (**Fig. 4C**), we found that cells in the core fell into a degenerated stress case (**Fig. S9**), where $\sigma_1$ and $\sigma_2$ were of similar magnitude. Accordingly, the maximal shear stress in the $\sigma_1$-$\sigma_2$ plane, $\tau_{12,max} = \sigma_1 - \sigma_2$, is relatively low, corresponding to the low driving force for cell rotation. At the same time, the high hydrostatic pressure at the core (**Fig. 2C**) corresponds to large shear resistance to cell rotation. Thus, the low shear stress and high resistance render cells at the core kinetically trapped. In this case, cells tend to avoid the most compressive ($\sigma_3$) direction but randomly orient in the $\sigma_1$-$\sigma_2$ plane, displaying a disordered phase. As shown in **Figs. 4C** and **4D**, cell ordering in the G-I biofilms can be divided into the outer bipolar region where cell highly aligns with the minimal compression, and the inner flat region where cell orientations avoid the maximal compression ($z$-direction).

Cell rotation relaxes the compressive stress, placing cells in a lower strain energy state. However, even though the cells near the interface undergo large rotations and align with their first principal stress, significant shear stress of the aligned cells remains in that region. This suggests that shear stress not only acts as the driving force for cell rotation and alignment, but also stabilizes the orientational pattern. Indeed, in the absence of shear stress, i.e., a planarly isotropic stress state ($\sigma_1 = \sigma_2$), there is no unique first-principal stress direction to align, and cells tend to randomly orient. Therefore, even though cell rotation tends to relax shear stress level, stress anisotropy is still necessary for keeping local cell alignments, or perturbations would drive cells back to the randomly oriented state.

To further interrogate the mechanical origin of cell ordering and the impact of shear stress, we analyzed the spatiotemporal evolution of cell ordering and stress anisotropy. We quantified stress anisotropy by the relative distance of the three principal stresses, $\alpha_\sigma = \frac{\sigma_1 - \sigma_2}{\sigma_1 - \sigma_3}$. As shown in **Fig. S9**, depending on $\alpha_\sigma$, a cell can exist in one of three possible stress states: $\alpha_\sigma \approx 1$ ($\sigma_1 > \sigma_2 > \sigma_3$), $\alpha_\sigma \approx 0$ ($\sigma_1 \approx \sigma_2 > \sigma_3$ or $\sigma_1 \approx \sigma_2 \approx \sigma_3$), and all other intermediate values $0 < \alpha_\sigma < 1$. When $\alpha_\sigma = 0$, the stress state is 2D or 3D isotropic stress state where the minimal compression direction is not uniquely defined. Following this definition, we find a strong spatiotemporal correlation between cell alignment and the stress anisotropy $\alpha_\sigma$, which suggests that stress anisotropy serves as a critical condition for cell orientational ordering (**Fig. 4C**, **4E**, **Fig. S6**).

**Growth stress drives cell reorientation with a delay due to shear stress relaxation**

To further illustrate how growth-induced stress drives cell alignment, we designed a set of numerical experiments that artificially manipulate the growth process. As shown in **Fig. 5A**, compared with the control numerical experiments with normal cell division and elongation, blocking either cell division (elongation only) or cell elongation (division only) markedly lowers cell alignment. Completely blocking cell proliferation (no division and elongation) results in the lowest level of cell alignment. Noticeably, blocking cell elongation has a larger impact on cell reorientation than blocking division, indicating that cell elongation provides the primary driving force for the cell reorientation. On the other hand, artificially imposing additional in-plane compression (20kPa) by laterally squeezing the surrounding hydrogels (Supplementary Materials) to the growing biofilm promotes cell alignment. These numerical experiments confirm that the growth stress contributes to the orientational ordering of the growing biofilms and shows the dual role of cell growth: it generates stress in biofilms, and at the same time is required for cell reorganization.

We further notice that the orientational cell ordering driven by growth stress is not spontaneous but manifests a time delay from growth-induced shear forcing. This delay is clearly seen from the time-lagged cross correlation between the shear stress and orientational ordering of cells near the biofilm-gel interface (**Fig. 5B, 5C**). Under normal growth conditions (**Fig. 5B**), the alignment near the gel-cell interface within the first half of $T_{\text{double}}$ (doubling time) hardly changes, while the equivalent shear stress markedly increases. This leads to appreciable alignment in the second half period of $T_{\text{double}}$. In the numerical experiment with artificially imposed in-plane compression (**Fig. 5C)**, the applied stress anisotropy is spontaneously transmitted to most of the cells, aligning the cells along the out-of-plane direction $\hat{n}_z$. However, the change in the stress state does not cause spontaneous cell reorientation. Rather, only after 10 minutes do the cells in the core start to reorient to $\hat{n}_z$ and the reorientation process continues for roughly 20 minutes. Given the average doubling time $T_{\text{double}} = 45\text{mins}$, this delay is close to half of $T_{\text{double}}$, which is associated with shear stress relaxation in orientational ordering.

**Mechanical control of self-patterning of growing biofilms**

The stress-driven morphodynamics sheds light on the control of cell orientational packing in growing biofilms. To further explore this direction, based on the wildtype simulation **(Fig. 6A)** we created three additional "mutants" in our simulations for the G-II biofilms, by varying the biofilm-substrate adhesion and biofilm-gel adhesion. Knocking down the biofilm-substrate adhesion diminishes the substrate friction, which eliminates the aster ordering at the basal plane (*21*) (**Fig. 6B**). Blocking off the biofilm-gel adhesion inhibits the stress transmission from gel to the biofilm (*20*) and diminishes the shear stress, which suppresses the bipolar nematic order near the interface (**Fig. 6C**). Deleting adhesion at both interfaces leads to a superimposed effect where the bacterial cells form an isotropic colony (**Fig. 6D)**. In addition, as gel stiffness represents the extent of mechanical confinement and dictates the growth stress profiles, spatially patterning the gel stiffness, such as a soft-core-rigid-rim structure (**Fig. 6E**, Supplementary Information), alters the biofilm cell ordering and morphology. Specifically, more verticalized cells appear in the biofilm core due to weaker confinement in the vertical direction, and the bipolar alignment at the cell-gel interface becomes discontinuous at the soft-to-rigid transition region. Overall, the biofilm exhibits a curvature difference between the soft and rigid regions. This may open a new dimension for controlling the orientational packing of biofilms.

**Discussion**

Environmental confinement is increasingly seen as an important determinant of bacterial biofilm development (*39-43*). Here we show that biofilm growth under hydrogel confinement induces spatiotemporally heterogeneous stress that contributes to biofilm morphological evolution and internal orientational cell ordering. While hydrostatic stress drives viscous-to-elastic transition and promotes random cell ordering, shear stress stemming from the confinement boundaries drives cell rotation and promotes nematic order. The nematic order thus ties to the spatiotemporal stress anisotropy in biofilm development. The stress-regulated morphodynamics is distinctly different from the bacterial cell collectives in confinement-free environments (*44*). Our findings provide new perspectives on how mechanical stresses spatiotemporally shape the long-range organization in growing 3D biofilms and reciprocally how living active systems respond to the surrounding mechanical environments. The stress-regulated orientational ordering may be relevant in other living active collectives and could underpin their long-term development and fitness.

The distinct growth pattern of biofilms under physical confinement suggests possible strategies to control bacterial material architecture with engineered stress anisotropy. Indeed, it has been demonstrated that tuning matrix viscoelasticity can achieve control of the spatiotemporal order of bacterial active matter and drive transitions from bacterial turbulence to unidirectional and oscillatory giant vortex (*24*). In our study, simple physical confinement results in complex spatiotemporal heterogeneities in mechanical stress, material phase, and orientational order, which are intimately coupled. The self-patterned architecture underlies percolation channels for the delivery of nutrients and signaling molecules essential to the fitness and further development of the biofilms (*9, 29*), which may also function as scaffolds for wound healing (*45-47*) and tissue regeneration (*48*). Our study thus inspires mechano-lithography of biofilms and other living active matter systems and paves the way to the development of a new class of adaptive self-driven devices and materials (*7*).

Despite the success of our agent-based model in predicting the stress-regulated orientational ordering in growing biofilms, extensions are yet to be made to recapitulate the mechanobiological feedback important for biofilm development. It has been shown that mechanical stress may upregulate stress-dependent factors at all stages of biofilm growth (*2, 14, 15*). For example, mechanical stress can induce secretion of the extracellular matrix such as exopolysaccharide (EPS) (*49, 50*), which in turn modulates the local mechanical properties of the growing films and feedbacks to stress generation. Stress landscape is spatially correlated with the accessibility of nutrients and antibiotics (*51*), and hence cell proliferation rate, which also feedbacks to stress generation. Furthermore, intrinsic heterogeneity of bacterial cells may also yield differences in growth stress generation and their responses to mechanical stress (*52*). A refined agent-based model is critically needed to account for these feedback loops in biofilm development. Such progress is essential for identifying new mechanobiological strategies towards controlling biofilm growth under relevant physiological conditions, which remains one of the foremost frontiers in bacterial biofilm development.

**Materials and Methods**

Agent-based simulations

We build our agent-based model based on the model developed by Beroz et al (*17*). Individual cells are treated as elongating and dividing spherocylinders with length $L(t)$ and radius $R$. We assume the growth of each cell follows exponential rule $\frac{dV}{dt} = \gamma V$ and assign a certain level of randomness $\gamma \sim \mathrm{N}(\gamma_0, 0.2\gamma_0)$ to the growth rate $\gamma$ across the biofilm. The interaction between cells is described by linear elastic Herzian contact mechanics (*53*) without cell-cell adhesion, and A modified JKR model (*25, 26*) is applied to represent the cell-substrate and cell-gel contact/adhesion. For individual cells, we also include environmental viscosity from the extracellular matrix environment and surface friction between the substrate, keeping cell dynamics always in the overdamped regime. The hydrogel confinement is modeled as a homogeneous, isotropic, and viscoelastic material using a coarse-grained approach. The gel particles are assumed as spheres with a radius $R_{gel}$, and the interactions are designed as a harmonic pairwise potential and a three-body potential related to bond angles. Gel-substrate interactions are similarly treated as the modified JKR model, and we considered water viscosity to stabilize the gel particle system. The full details of agent-based simulations are contained in Supplementary Information.

**Acknowledgements:**

We thank Dr. R. Alert for his helpful discussions.

**Funding:**

Research reported in this publication was supported by the National Institute of General Medical Sciences of the National Institutes of Health under Award Number DP2GM146253. J.Y. holds a Career Award at the Scientific Interface from the Burroughs Welcome Fund (1015763.02).

**Author Contributions:**

C.L. and S.Z. developed the agent-based simulations. J.N., Q.Z., J.Y. designed and performed the experiments. C.L, J.N., L.F., Q.Z., J.Y., and S.Z. analyzed data. C.L., J.N., L.F., J.Y., and S.Z. wrote the paper.

**Competing interests:**

The authors declare that they have no competing interests.

**Data and materials availability:**

The code for agent-based simulation is available online: https://github.com/LAMMPS-Agent/LAMMPS-Agent/tree/1.0. Other data are available upon request.

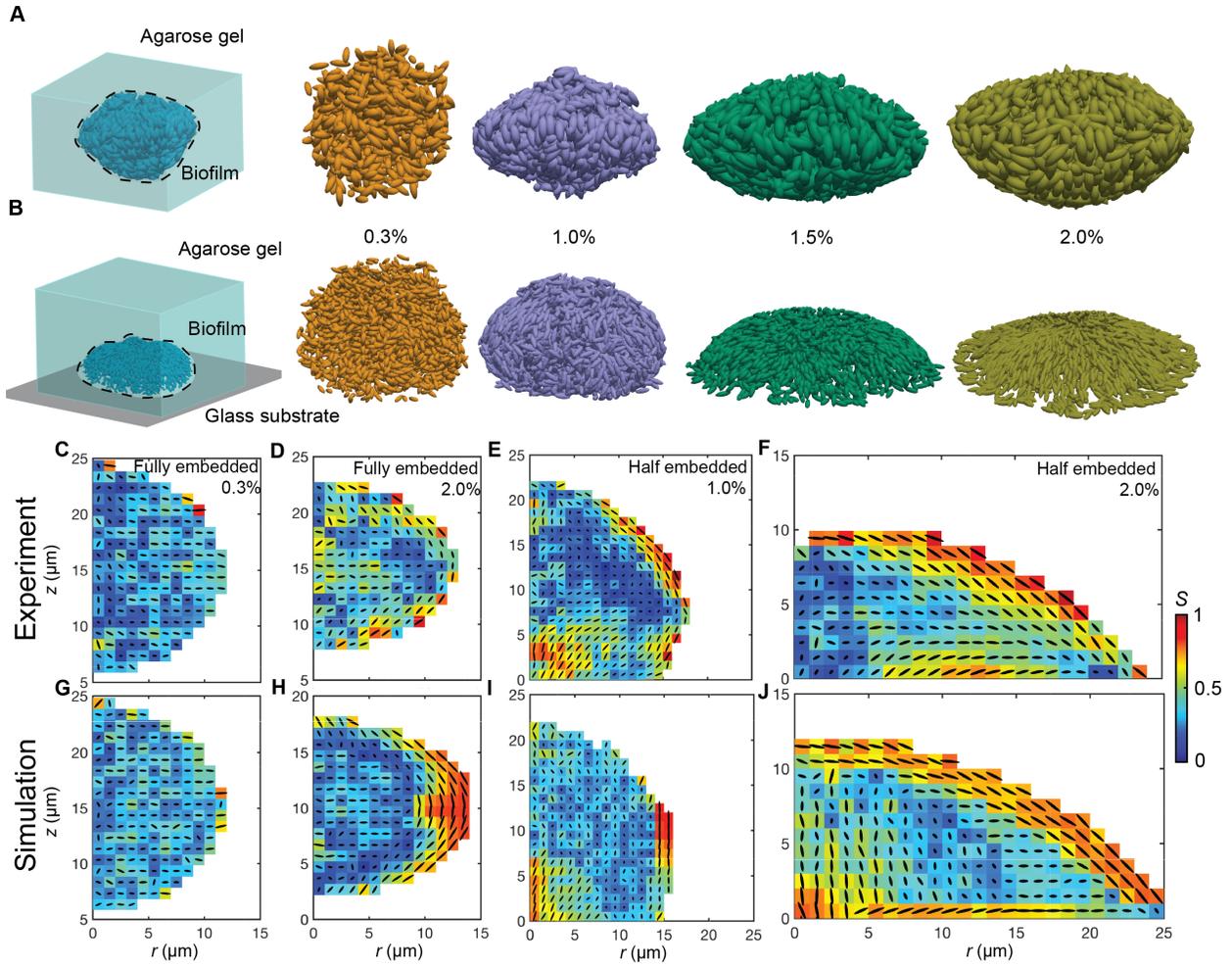

**Fig. 1. Confinement-dependent growth morphology and cell alignment of biofilms. A, B** 3D morphology and cell ordering of G-I (A) and G-II (B) biofilms. Shown from left to right are reconstructed experimental biofilm morphologies embedded in agarose gels with a gel concentration of 0.3%, 1.0%, 1.5% and 2.0%, respectively. Scale bar: 5µm. **C-F**. Reconstructed cell ordering in the experimental biofilms, color-coded by the local nematic order parameter $S$. Cell position and orientation are azimuthally averaged in bins of $1\mu m \times 1\mu m$. Black ovals represent the azimuthal projection of the averaged orientation vector $\hat{\boldsymbol{n}}_c$. **G-J**. Simulated biofilm morphology and cell order by agent-based modeling, corresponding to **C-F**.

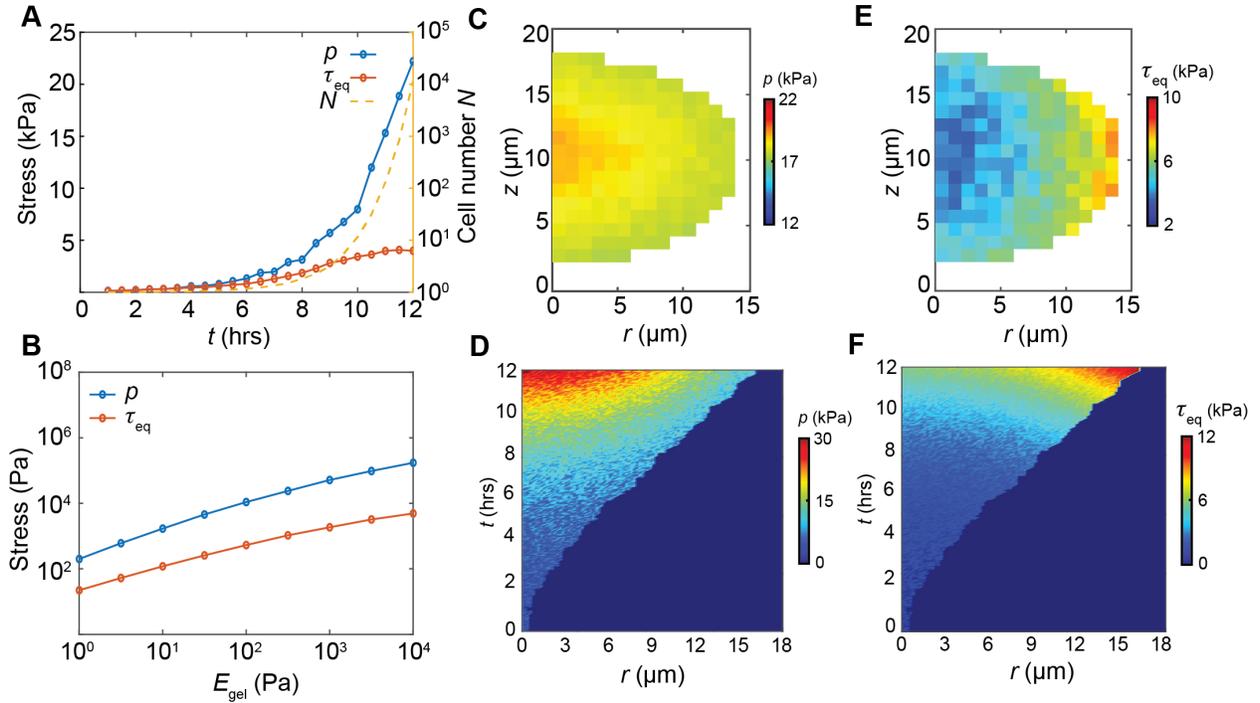

**Fig. 2. Spatiotemporal stress evolution in G-I biofilms. A.** Time series of spatially averaged pressure $p$, equivalent shear stress $\tau_{eq}$, and cell number $N$ of simulated biofilms with $E_{gel} = 100$ kPa (corresponding to an agarose concentration of 2% in the experiment). **B.** Average hydrostatic and equivalent shear stresses monotonically increase with gel stiffness. Each data point is averaged across 5 biofilms for similar cell numbers (~4500 cells). **C.** Spatial distribution of hydrostatic pressure $p$ of the biofilm (growth time ~11 hrs.). The data is azimuthally averaged in bins of 1μm × 1μm. **D.** A kymograph of hydrostatic pressure $p$, where $r$ measures the long axis distance from the center to the gel-cell interface of the biofilm. **E-F.** The spatial distribution and kymograph of the equivalent shear stress, showing a shear boundary layer.

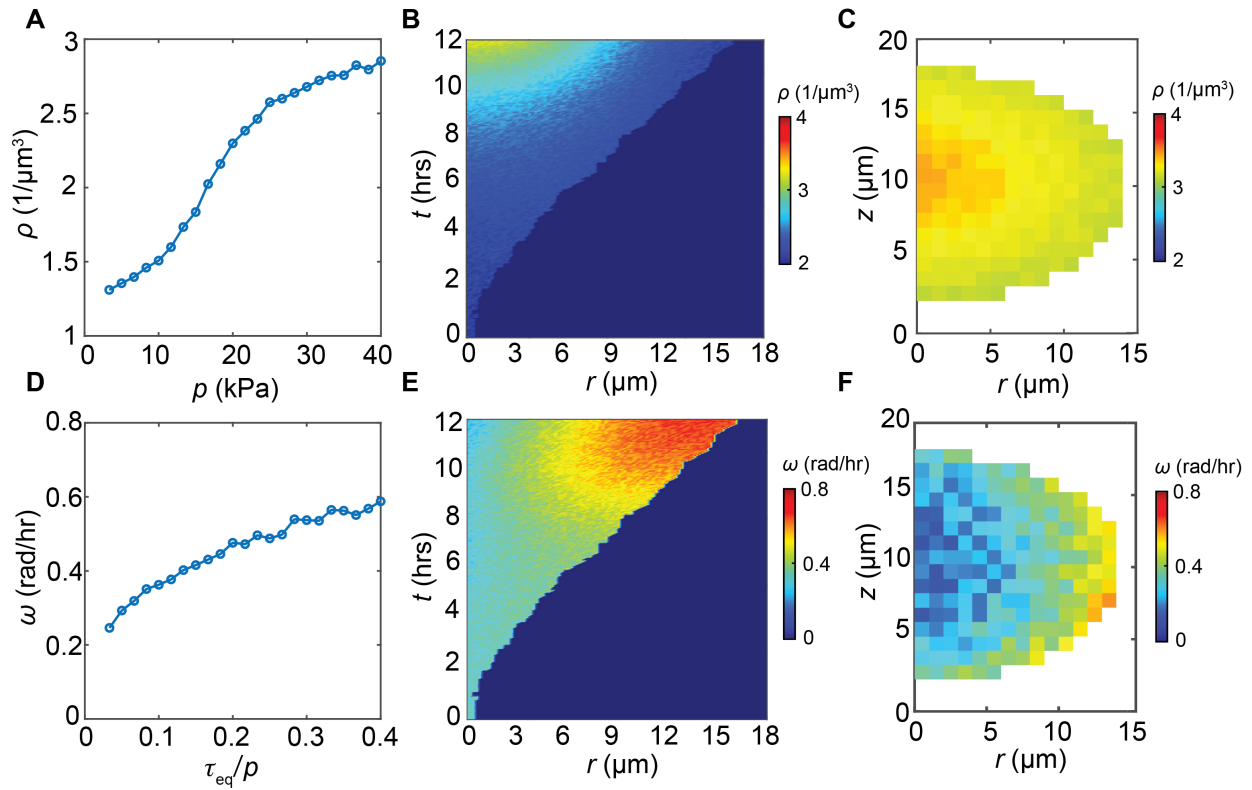

**Fig. 3. Viscous-to-elastic transition and heterogeneous cell ordering in confined biofilms**. **A.** Biofilm density increases with hydrostatic pressure. **B.** A kymograph of cell density evolution, where $r$ measures the distance from the center of the biofilm. **C.** Spatial distribution of biofilm density in mature biofilms. **D.** Shear stress promotes and hydrostatic pressure resists cell rotation. The cell rotation speed depends on the shear stress normalized by hydrostatic pressure. **E-F.** kymograph (**E**) and spatial distribution (**F**) of rotation speed show that the bacterial cell reorientation occurs at the gel-cell interface but is inhibited at the core of the biofilms.

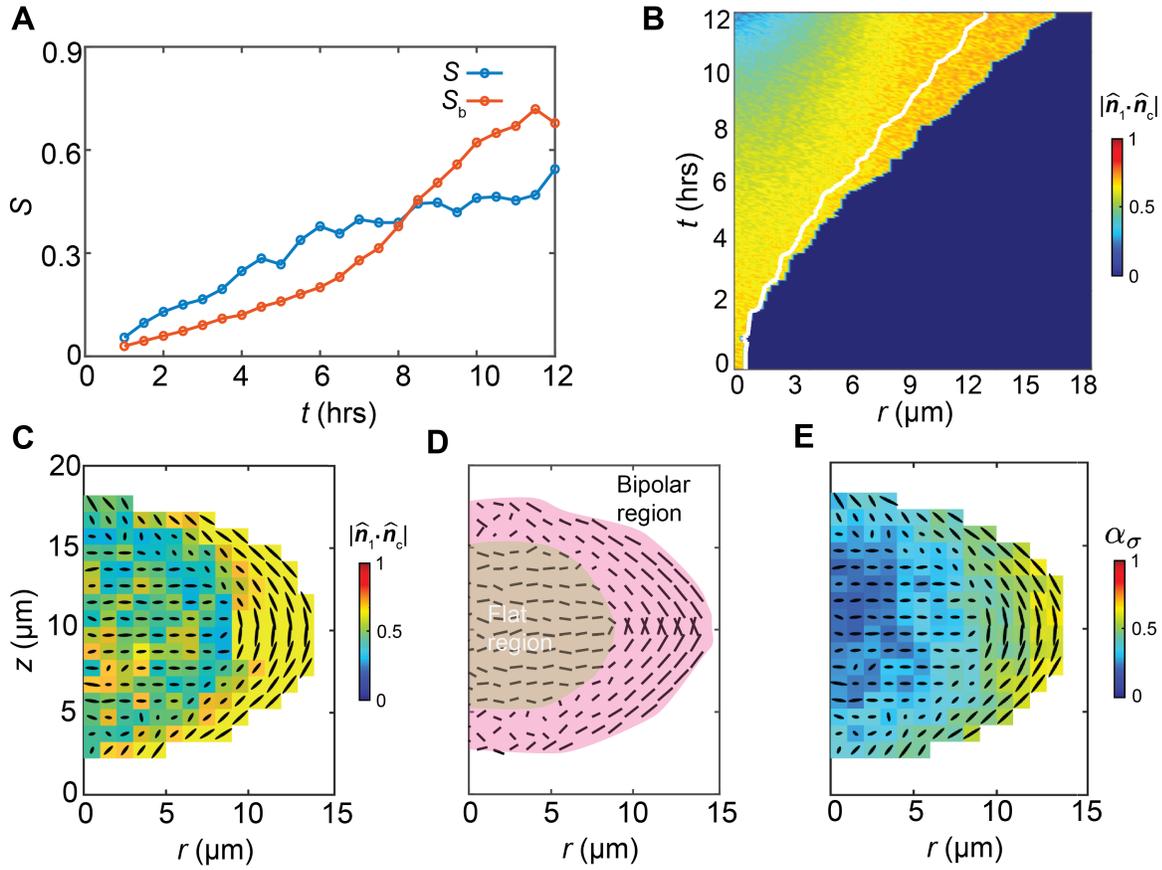

**Fig. 4. Stress anisotropy and cell ordering of G-I biofilms. A.** Time evolution of nematic ordering $S$ and the boundary bipolar ordering $S_b$. **B-C.** The kymograph (**B**) and spatial distribution (**C**) of the relative orientation parameter $\alpha$ show that cells tend to rotate to achieve bipolar alignment at the gel-cell interface. The white line in **B** separates the nematically ordered phase and the disordered phase. **D.** The spatial distribution of the direction of the first principal stress $\hat{n}_1$. **E.** The spatial distribution of the stress anisotropy parameter $\alpha_\sigma$ coincides with nematic cell order.

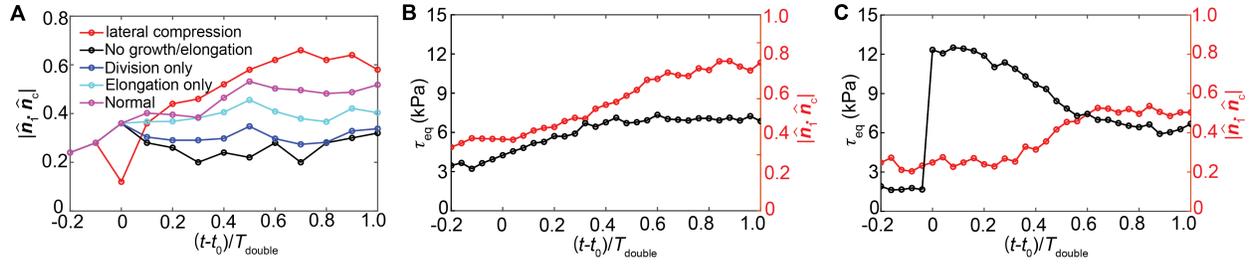

**Fig. 5. Kinetics of stress-induced cell reorientation. A.** The time evolution of the relative orientation α under different conditions of elongation and division. **B-C.** The time evolution of average equivalent shear stress $\tau_{eq}$ and the average z-component of cell directors, for normal growth **(B)** and lateral compression **(C)**. For clarity, in **(B)** we only consider the boundary aligned layer of thickness 5μm, to get rid of orientation randomness from the biofilm core.

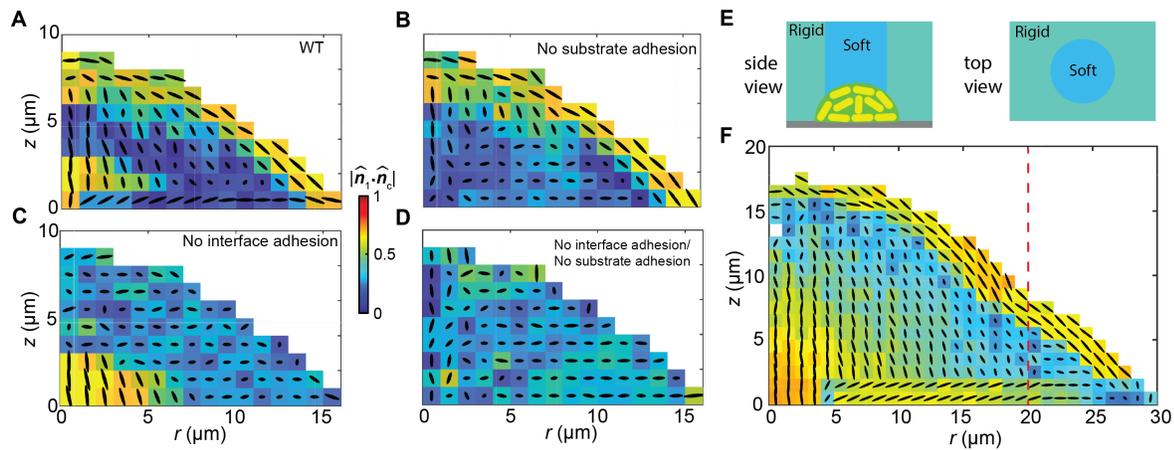

**Fig. 6. Mechanobiological control of orientational packing in biofilms. (A).** WT biofilm with normal substrate-biofilm adhesion and gel-biofilm adhesion **(B).** Mutant biofilms with the knocking-down of substrate-biofilm adhesion eliminates the radial ordering in the basal plane. **(C)** Mutant biofilms with the knocking-down biofilm-gel adhesion suppress the bipolar alignment at the interface. **(D)** Mutant biofilms with both adhesions knocked down results in isotropic cell packing for the entire biofilm. **(E)** Schematic illustration of the numerical experiment of gel stiffness patterning. The soft, cylindrical region is 20 μm in radius and 1/25 of the stiffness in the rigid region. **(F)** Spatially patterned gel stiffness modifies cell orientational alignment. Red dashed line represents the soft-rigid boundary of the gel.

# Supplementary Materials for

## Mechano-lithography: stress anisotropy drives nematic ordering in growing three-dimensional bacterial biofilms

Changhao Li, Japinder Nijjer, Luyi Feng, Qiuting Zhang, Jing Yan*, Sulin Zhang*

*Corresponding author. Email: suz10@psu.edu or jing.yan@yale.edu

**This PDF file includes:**

Supplementary Text
Figs. S1 to S9

**Supplementary Text**

Agent-based model (ABM)

Our agent-based model is built on the previous model developed by Beroz et al. (*1*) and others (*2, 3*), and has successfully applied to the differential growth and self-patterning problem of *V. cholorae* biofilms (*4*). As shown in **Fig. S1**, the biofilm is modeled as elongating and dividing spherocylindrical agents and the surrounding hydrogels are modeled as spherical agents. The details of agent particle geometries, interactions, governing equations, and parameter setting are introduced below.

*Single cell geometry*

For simplicity, we assume that the single spherocylinder represents the space occupied by a single cell and its surrounding extracellular matrix. Each spherocylinder can be determined by its length $L$ and radius $R$, with the volume $V = \frac{4}{3}\pi R^3 + \pi R^2 L$. For a single cell, we assume the cell length $L$ constantly grows with time, and the radius $R$ keeps the same, leading to the exponential volume growth $\frac{dV}{dt} = \gamma V$. To introduce randomness into the model, we assume the growth rate $\gamma$ follows the normal distribution $\gamma \sim N(\gamma_0, 0.2\gamma_0)$, where $\gamma_0$ is the average growth rate calibrated from experiments. In simulations, the continuous exponential growth is implemented as discrete length increments of $\Delta L = \gamma \left(\frac{4}{3}R + L\right)\Delta t$, where $\Delta t$ is the timesteps.

We model the cell division as the following: when the length of a mother cell reaches the division length $L_{\max}$, it is instantaneously replaced by two equal-sized daughter cells with initial length $L_0 = \frac{L_{\max}}{2} - R$. As shown in **Fig. S1**, $L_0$ is determined by the criteria where two daughter cells have the same total head-to-tail length as the mother cell. It follows, for a cell with initial length $L_0$ and growth rate $\gamma$, that the doubling time is $t_{\text{double}} = \frac{1}{\gamma}\log\left(\frac{10R+6L_0}{4R+3L_0}\right)$. In simulations, the division is implemented as the following: In a single timestep, if the length of a mother cell reaches $L_{\max}$, the length of this cell is altered to $L_0$, as the first daughter cell. The second daughter cell is generated by direct copy of the first; then their center positions are changed so that they occupy the same head-to-tail position as the mother cell without overlapping.

It should be noted that, we use a hard-core, soft-shell model to capture the mechanical property of the cell-matrix composite. We assume the spherocylindrical region of a single cell can be divided into two regions with different contact stiffness $E$. The outer region represents the soft extracellular matrix ($E_{\text{mat}} \sim 300$ Pa) (*5*), and the inner region represents the rigid bacteria cell ($E_{\text{cell}} \sim 50$ kPa) (6). We denote the radius of rigid cell (inner region) by $R_c$ to differentiate it with the radius of cell-matrix composite $R$, and all the word "cell" represents "cell-matrix composite" in the following descriptions, unless any special declaration.

*Cell-cell repulsion*

For cell-cell interactions, we neglect adhesion and friction between cells, but only consider elastic contacts. We apply linear elastic Herzian contact theory (7) to quantify the repulsive contact force on cell $i$ by cell $j$, written as

$$\boldsymbol{F}_{\text{cell-cell},ij} = \begin{cases} -\dfrac{5}{2} E_0 R^{1/2} \delta_{ij}^{3/2} \hat{\mathbf{e}}_{ij}, \delta_{ij} < R - R_c \\ -\dfrac{5}{2} (E_0 R^{1/2}(R - R_c)^{3/2} + E_c R^{1/2}(\delta_{ij} - R + R_c)^{3/2}) \hat{\mathbf{e}}_{ij}, \delta_{ij} > R - R_c \end{cases} \quad (1)$$

where $E_0$ and $E_c$ denotes the effective contact stiffness of extracellular matrix and rigid cell respectively, $R_c$ is the radius of center rigid cell, $\delta_{ij}$ is the overlapping distance, and $\hat{\mathbf{e}}_{ij}$ denotes the unit vector normalized from the distance vector $\boldsymbol{d}$, defined as the smallest distance between two cell centerlines. The overlapping distance $\delta_{ij}$ is given by $\delta_{ij} = 2R - |\boldsymbol{d}|$. Correspondingly, the moment of contact force $\boldsymbol{F}_{\text{cell-cell},ij}$ about cell center is given by

$$\boldsymbol{M}_{\text{cell-cell},ij} = s_r \hat{\boldsymbol{n}}_i \times \boldsymbol{F}_{\text{cell-cell},ij} \quad (2)$$

where $\hat{\boldsymbol{n}}_i$ is the unit vector denoting cell orientation (from cell center to the contact point) and $s_r$ is the parametric coordinate of the contact point along the center line of the cell.

*Cell-gel interactions*

On the interface between biofilm and the surrounding hydrogels, we apply a JKR-type model to capture the elastic contact and adhesion between cell agents $i$ and coarse-grained gel particles $j$. For the elastic contact, we use similar linear elastic Herzian contact interaction as Eq. (1), written as

$$\boldsymbol{F}_{\text{cell-gel},ij}^{rep} = \begin{cases} -\dfrac{5}{2} E_0 R_{\text{eq}}^{1/2} \delta_{ij}^{3/2} \hat{\mathbf{e}}_{ij}, \delta_{ij} < R - R_c \\ -\dfrac{5}{2}(E_0 R_{\text{eq}}^{1/2}(R - R_c)^{3/2} + E_c R_{\text{eq}}^{1/2}(\delta_{ij} - R + R_c)^{3/2}) \hat{\mathbf{e}}_{ij}, \delta_{ij} > R - R_c \end{cases} \quad (3)$$

where $R_{eq} = \dfrac{2R_{\text{gel}}R}{R + R_{\text{gel}}}$ is the equivalent radius of contact, $R_{\text{gel}}$ is the radius of coarse-grained gel particles, the overlapping $\delta_{ij} = R + R_{\text{gel}} - |\boldsymbol{d}|$ is determined by similar methods considering the minimal distance between the center of the gel particle to the cell center line.

We assume the adhesion force is proportional to $\gamma_{\text{cell-gel}}$, the energy release per unit area given by $\gamma_{\text{cell-gel}} = \gamma_{\text{cell}} + \gamma_{\text{gel}} - 2\gamma^*$, the surface energy of cell and gel minus double interface energy $\gamma^*$. Naturally, the adhesion force is also proportional to the contact area, which gives

$$\boldsymbol{F}^{adh}_{\text{cell-gel},ij} = \pi a^2 \gamma_{\text{cell-gel}} \hat{\boldsymbol{e}}_{ij} \tag{4}$$

where $a$ is the equivalent radius of contact area, given by the simplified relation $a = \sqrt{\delta_{ij} R_{\text{eq}}}$. For simplicity, our model neglects the cohesion-decohesion asymmetric behavior in the original JKR model, as the decohesion behavior rarely happens on the continuously expanding biofilm-gel interface.

Similarly, the moment about the cell center, for the repulsive cell-gel contact force $\boldsymbol{F}^{rep}_{\text{cell-gel},ij}$ and the adhesion $\boldsymbol{F}^{adh}_{\text{cell-gel},ij}$, can be given by

$$\boldsymbol{M}^{rep}_{\text{cell-gel},ij} = s_r \hat{\boldsymbol{n}}_i \times \boldsymbol{F}^{rep}_{\text{cell-gel},ij} \tag{5}$$

and

$$\boldsymbol{M}^{adh}_{\text{cell-gel},ij} = s_r \hat{\boldsymbol{n}}_i \times \boldsymbol{F}^{adh}_{\text{cell-gel},ij} \tag{6}$$

*Gel-gel interactions*

We treat the surrounding agarose gel as soft viscoelastic material, using coarse-grained modeling to address its mechanical behavior. The basic element of the coarse-grained model is spherical particles of radius $R_{\text{gel}}$ with harmonic interactions. We set the pairwise interaction energy between the gel particles as $E_{\text{gel},2} = \Sigma_{ij} \frac{k_r}{2}(\xi_{ij} - \xi_0)^2$, with the cut-off radius $R^c_{\text{gel},2}$, where $\xi_{ij}$ is the distance between particle $i$ and $j$, $\xi_0$ is the equilibrium distance, and $k_r$ is the spring constant. To capture the gel shear modulus, we also introduce a three-body interaction energy where $E_{\text{gel},3} = \Sigma_{ijk} \frac{k_\zeta}{2}(\zeta_{ijk} - \zeta_0)^2$, with the cut-off radius $R^c_{\text{gel},3}$, where $\zeta_{ijk}$ is the bond angle formed by particle $i, j$, and $k$. We also considered the normalized stokes viscosity of gel particles to address the viscoelasticity behavior also stabilize the gel system, given by $\boldsymbol{F}_{\text{stokes},i} = -\eta_{gel} \boldsymbol{u}_i$, where $\eta_{gel}$ is the normalized viscosity coefficient and $\boldsymbol{u}_i$ is the velocity vector of the gel particle $i$.

*Cell-to-substrate interactions*

Considering the glass substrate in experiments has significantly larger Young's modulus compared with cells and gels, we model the substrate as an infinite rigid, two-dimensional plane located at $z = 0$. Similarly, we apply linear elastic Herzian contact model to represent the repulsion between cells and substrate. On the other hand, we assume the cell-substrate adhesion is related to the equivalent contact area between cell and substrate, by Derjaguin approximation (8).

For the cell-substrate repulsive contact, we use a generalized Herzian contact formula to account for the cell orientation dependent contact energy. Similar to Eq. (1) and (2), the elastic contact energy is given by $E_{el,i} = E_0 R^{1/2} \delta_i^{5/2}$, where $\delta_i$ is the equivalent penetration depth which depends on the average penetration depth and the cell-substrate relative angle, given by the explicit formula:

$$\delta_i^{5/2} = \int_{-L/2}^{L/2} \left[ R^{1/2} |\hat{n}_{\parallel,i}|^2 \delta^2(s) + \frac{4}{3}\left(1 - |\hat{n}_{\parallel,i}|^2\right) \delta^{3/2}(s) \right] ds \quad (7)$$

where $\hat{n}_{\parallel,i}$ is the normalized projection of the $i$th cell director onto the substrate ($z = 0$). The overlap function $\delta(s)$ denotes the overlapping distance between the cell and the substrate at the local cell-body coordinate $-L/2 \leq s \leq L/2$. Then, the net force $\boldsymbol{F}_{el,i}$ and moment $\boldsymbol{M}_{el,i}$ from the cell-substrate elastic repulsion can be given by

$$\boldsymbol{F}_{el,i} = 2E_0 R^{1/2} \int_{L/2}^{-L/2} \hat{\boldsymbol{z}} \left[ R^{1/2} |\hat{n}_{\parallel,i}|^2 \delta(s) + \left(1 - |\hat{n}_{\parallel,i}|^2\right) \delta^{1/2}(s) \right] ds \quad (8)$$

$$\boldsymbol{M}_{el,i} = 2E_0 R^{1/2} \int_{L/2}^{-L/2} [s\hat{\boldsymbol{n}}_i \times \hat{\boldsymbol{z}}] \left[ R^{1/2} |\hat{n}_{\parallel,i}|^2 \delta(s) + \left(1 - |\hat{n}_{\parallel,i}|^2\right) \delta^{1/2}(s) \right] ds \quad (9)$$

where $\hat{\boldsymbol{z}}$ is the unit vector perpendicular to the substrate.

Similar to Eq. (4), we assume the cell-substrate adhesion energy by the form of $E_{ad,i} = -\Sigma_0 A_i$, where $\Sigma_0$ is the adhesion energy density and $A_i$ is the equivalent contact area between cell $i$ and the substrate. The equivalent contact area is given by

$$A_i = \int_{-L/2}^{L/2} a(s)ds = \int_{-L/2}^{L/2} \left[ R^{1/2} |\hat{n}_{\parallel,i}|^2 \delta^{1/2}(s) + \pi R \left(1 - |\hat{n}_{\parallel,i}|^2\right) H(\delta(s)) \right] ds \quad (10)$$

where $H(\cdot)$ is the Heaviside step function. Thus, the net adhesive force $\boldsymbol{F}_{ad,i}$ and moment $\boldsymbol{M}_{ad,i}$ are:

$$\boldsymbol{F}_{ad,i} = -\Sigma_0 \int_{-L/2}^{L/2} \hat{\boldsymbol{z}} \left[ \frac{1}{2} R^{1/2} |\hat{n}_{\parallel,i}|^2 \delta^{-1/2}(s) \right] ds - \hat{\boldsymbol{z}} \Sigma_0 \pi R \left(1 - |\hat{n}_{\parallel,i}|^2\right) \quad (11)$$

$$M_{\text{ad},i} = -\Sigma_0 \int_{-L/2}^{L/2} [s\widehat{\boldsymbol{n}}_i \times \widehat{\boldsymbol{z}}] \left[\frac{1}{2}R^{1/2}|\widehat{\boldsymbol{n}}_{\parallel,i}|^2 \delta^{-1/2}(s)\right] ds \qquad (12)$$
$$- [s_0\widehat{\boldsymbol{n}} \times \widehat{\boldsymbol{z}}]\Sigma_0 \pi R \left(1 - |\widehat{\boldsymbol{n}}_{\parallel,i}|^2\right)$$

where $s_0$ denotes the cell-body coordinate such that $\delta(s_0) = 0$. Namely, $\delta(s_0) = 0$ gives the point where a cell detaches from the substrate.

*Viscosity of cells*

We consider two sources of viscosity of cells: a bulk viscous force due to the friction from extracellular matrix environment and a surface viscous force due to the substrate. The environmental viscous force and moment are given by Stoke's law,

$$\boldsymbol{F}_{\text{stokes},i} = -\eta_0 \boldsymbol{u}_i \qquad (13)$$

$$\boldsymbol{M}_{\text{stokes},i} = -\eta_0 \int_{L/2}^{L/2} s\widehat{\boldsymbol{n}}_i \times (\boldsymbol{\omega}_i \times s\widehat{\boldsymbol{n}}_i) ds = -\frac{\eta_0}{12}\boldsymbol{\omega}_i L^3 \qquad (14)$$

Where $\eta_0$ is the normalized environmental viscosity, $\boldsymbol{u}_i$ is the velocity of the center of mass, and $\boldsymbol{\omega}_i$ is the angular velocity. The substrate viscous force and moment are taken to be of the form

$$\boldsymbol{F}_{\text{surface},i} = -\int_{-L/2}^{L/2} \frac{\eta_1 a(s)}{R}[\boldsymbol{u}_i(s) - (\boldsymbol{u}_i(s)\cdot\widehat{\boldsymbol{z}})\widehat{\boldsymbol{z}}]ds \qquad (15)$$

$$\boldsymbol{M}_{\text{surface},i} = -\int_{-L/2}^{L/2} \frac{\eta_1 a(s)}{R} s\widehat{\boldsymbol{n}}_i \times [\boldsymbol{u}_i(s) - (\boldsymbol{u}_i(s)\cdot\widehat{\boldsymbol{z}})\widehat{\boldsymbol{z}}]ds \qquad (16)$$

where $\eta_1$ is the viscous coefficient along the substrate.

*Interactions between the gel particles and the substrate*

We use two types of interactions to mimic the experimental condition where the gel is adhered to the substrate. The first type of interaction is gel-substrate elastic contacts. Here we again apply the linear elastic Herzian contact theory between a sphere and a flat rigid surface, and the elastic contact energy is given by $E_{\text{gel-surface},i} = E_1 R_{\text{gel}}^{1/2} \delta_i^{5/2}$, where $E_1$ is the contact stiffness between gel particles and the substrate and $\delta_i$ is the overlap between the gel particle and the substrate. The second type is the gel-substrate adhesion, serving as the energy barrier for the experimentally observed delamination on the gel-substrate interfaces. Similarly, we take the adhesion energy as $E_{\text{ad,gel},i} = -\Sigma_1 A_{\text{gel},i}$, where $\Sigma_1$ is the adhesion energy density and the equivalent contact area is given by $A_{\text{gel},i} = \pi R_{\text{gel}} \delta_i$.

*Equations of motion*

The equations of motion for each cell are given by Newton's rigid body dynamics:

$$\begin{pmatrix} \boldsymbol{F}_{\text{net},i} \\ \boldsymbol{M}_{\text{net},i} \end{pmatrix} = \begin{bmatrix} m & 0 \\ 0 & \mathbf{I}_i \end{bmatrix} \begin{pmatrix} \dot{\boldsymbol{u}}_i \\ \dot{\boldsymbol{\omega}}_i \end{pmatrix} + \begin{pmatrix} 0 \\ \boldsymbol{\omega}_i \times \mathbf{I}_i \boldsymbol{\omega}_i \end{pmatrix} \tag{17}$$

where $\boldsymbol{F}_{\text{net},i}$ and $\boldsymbol{M}_{\text{net},i}$ are the total force and moment vector, and $\mathbf{I}$ is the moment of inertia. All of the variables are expressed in the body-fixed coordinate system, then transformed into the global coordinate system. We add a small random noise to the net force and moment vectors of the cells at every timestep ($10^{-7} E_0 R^2$ for forces and $10^{-7} E_0 R^3$ for moments), to represent the environmental fluctuations required for breaking the symmetry.

We only consider the three translational degrees of freedom for gel particles, neglecting the rotational degrees of freedom. Therefore, the equations of motion are given by Newton's second law $\mathbf{F}_{\text{tot},i} = m \dot{\boldsymbol{u}}_i$, where $\mathbf{F}_{\text{tot},i}$ is the net force and $\dot{\boldsymbol{u}}_i$ is the acceleration. To prepare the initial amorphous stress-free geometry, we begin with a body-centered cubic crystalline geometry with lattice parameter $a$, where $a = 1.15 R_{\text{gel}}$. Subsequently, we assigned the system with an initial temperature of $300K$ and annealed it (using NVT thermostat) until it reached a final configuration that is amorphous and stress-free (spatial averaged residual pressure smaller than 0.01 kPa).

*Choice of parameters*

The cell radius $R$ and the division (maximum) length $L_{\max}$: we set $R = 0.8$ μm and $L_{\max} = 3.6$ μm to match the experimentally measured mean radius and division length.

The hard-core stiffness $E_c$, hard-core radius $R_c$, and soft-shell stiffness $E_0$: we set hard-core stiffness $E_c = 30$kPa to match the reported experimental measurement (8 to 47kPa) (*6*), while the soft-shell stiffness $E_0 = 300$Pa according to the previous rheology experiments (*1*). We set the hard-core radius as $R_c = 0.5$μm.

viscosity coefficients $\eta_0, \eta_1, \eta_{gel}$: We take $\eta_0 = 2 \times 10^{-6}$ Pa·s·m, and $\eta_1 = 2 \times 10^5$ Pa·s to match the previous rheology experiments. Taken the water viscosity $\mu_W = 8.9 \times 10^{-4}$ Pa·s, $\eta_{gel}$ is calculated as $\eta_{gel} = 6\pi\mu_W R_{gel} \approx 2 \times 10^{-8}$ Pa·s·m. Interestingly, we found $\eta_1$ and the gel stiffness jointly control the biofilm morphology, reported in the reference.

The spring constant $k_r$ and equilibrium length $\xi_0$: The Young's modulus of the agent-based gel system is given by (9) $Y = \frac{1}{V}\frac{\partial^2(E_{gel,2}+E_{gel,3})}{\partial \epsilon^2} \simeq \frac{k_r}{2\xi_0}$, under the condition $k_\zeta \ll k_r$. Generally, a smaller $\xi_0$ leads to a denser gel system and better approximation to a continuum solid. Here we choose $\xi_0 = 0.6$ µm as a result of a trade-off between simulation quality and computational cost, as the simulation time is proportional to $\frac{1}{\xi_0^3}$. $k_r$ ranges from $1.2 \times 10^{-4}$ Nm$^{-1}$ to $1.2 \times 10^{-1}$ Nm$^{-1}$ corresponding to the Young's modulus $Y = 0.1$ kPa to $Y = 100$ kPa.

Radius of agent gel particle $R_{gel}$: in order to mimic the continuum constraints posed by the hydrogel in the experiment, $R_{gel}$ should be larger than $\xi_0$. On the other hand, $R_{gel}$ cannot be significantly larger than the cell radius $R$, as this will introduce unphysical contacts at the biofilm-gel interface. Taking both requirements into consideration, we choose $R_{gel} = 1.0$ µm, which is nearly double the equilibrium distance $\xi_0 = 0.6$ µm and we keep $R_{gel} \approx R$.

Gel-substrate contact stiffness $E_1$ and adhesion energy density $\Sigma_1$: we choose $E_1 = 5\ kPa$ and $\Sigma_1 = 5 \times 10^{-2}$ N·m$^{-1}$ for all gel stiffnesses.

Simulation settings and boundary conditions

In simulations for both G-I and G-II biofilms, the surrounding hydrogel is modeled as a homogeneous, isotropic and linear elastic material using the agent-based model elaborated above. For the G-II biofilm, we set the simulation domain as the cubic box with the size of $200 \times 200 \times 120$ µm$^3$, where the initial geometry is initialized with a single cell lying parallel to the substrate without initial velocity and acceleration, surrounded by gel particles filling the entire simulation domain. A small hemisphere around the seed cell is vacated to avoid initial overlap between cell and hydrogel particles. For the G-I biofilms, the simulation domain size is $200 \times 200 \times 200$ µm$^3$ due to the removal of the rigid substrate. Similarly, the initial seed cell is placed in the center of the cubic box, with a small spherical region vacated. For both kind of simulations, we fix a small number of hydrogel particles near the x-y boundaries to provide anchoring for the elastic deformation of the hydrogel; however, the boundaries are kept sufficiently far away from the biofilm to minimize any boundary effects.

Calculation of the stresses

To quantify stress distribution in our complex system containing active growing/dividing bacteria and passive hydrogels, we define the stress tensor using Virial expression (*10*), augmented by a contribution from ambient viscosity. The stress tensor is naturally separated in a contribution from interactions and a contribution from environmental viscosity, as $\boldsymbol{\sigma} = \boldsymbol{\sigma}^{\text{int}} + \boldsymbol{\sigma}^{\text{vis}}$, with

$$\sigma_i^{\text{int}} = \frac{1}{V}\Sigma_j \boldsymbol{r}_{ij} \otimes \boldsymbol{F}_{ij} \qquad (18)$$

where $\boldsymbol{F}_{ij}$ is the summation of all particle-particle interactions between particle $i$ and $j$, and $\boldsymbol{r}_{ij}$ is the distance vector between particle $i$ and $j$, and $V$ is the cell volume. The viscosity part of Virial stress is given by

$$\sigma_i^{vis} = \frac{1}{V}\Sigma \boldsymbol{r}_i \otimes \boldsymbol{F}_i \qquad (19)$$

where $\boldsymbol{F}_i$ is any viscous force such as the ambient viscous force in Eq. (13) and the substrate viscous force in Eq. (15), $\boldsymbol{r}_i$ is the equivalent acting point of the given viscous force. The summation goes over all types of viscosity, and the definition of $\boldsymbol{\sigma}^{\text{int}}$ and $\boldsymbol{\sigma}^{\text{vis}}$ can be apply to both rod-shaped bacteria and sphere-shaped gel particles. Namely, for the translational ambient viscosity, the contribution to the viscosity stress can be written as $\sigma_i^{vis} = \frac{1}{V}\int_{-L/2}^{L/2} \boldsymbol{x} \otimes (-\eta\boldsymbol{v})\mathrm{d}\boldsymbol{x}$, where $\boldsymbol{v}$ is the velocity vector of the cell center of mass and $\boldsymbol{x}$ is the relative position vector to the cell center. Similarly, the contribution of rotational ambient viscosity can be given by $\sigma_i^{vis} = \frac{1}{2V}\int_{-L/2}^{L/2} \boldsymbol{x} \otimes (-\boldsymbol{\omega} \times \boldsymbol{x})\mathrm{d}\boldsymbol{x}$, where $\boldsymbol{\omega}$ is the angular velocity vector.

Calculation of cell ordering

We use the $\boldsymbol{Q}$-tensor model of liquid crystals (*11*) to quantify the local biofilm cell ordering. We calculate the per-cell traceless quantity $\boldsymbol{Q}_i = (3\hat{\boldsymbol{n}}_i \otimes \hat{\boldsymbol{n}}_i - \mathbf{I})/2$, where $i$ denotes the $i^{th}$ cell and $\mathbf{I}$ denotes the identity tensor. Compared with the cell director $\hat{\boldsymbol{n}}_i$, $\boldsymbol{Q}_i$ is head-tail symmetric given by $\boldsymbol{Q}_i(\hat{\boldsymbol{n}}_i) = \boldsymbol{Q}_i(-\hat{\boldsymbol{n}}_i)$. Considering the axisymmetric shape of biofilm, we use discretized bins under cylindrical coordinates $\Delta r = 1\mu\text{m}$, $\Delta z = 1\mu\text{m}$ and $\Delta\theta = \pi/4$ and average $\boldsymbol{Q}$ in each cylindrical bins generating the locally averaged order parameter $\boldsymbol{Q}(r_i, \theta_j, z_k)$, where $r_i, \theta_j, z_k$ denotes the bin numbered with $(i,j,k)$. The visualization of the azimuthally averaged $\boldsymbol{Q}$ is calculated by averaging the azimuthally projected order parameter $\boldsymbol{Q}_p = \boldsymbol{R}^T\boldsymbol{Q}\boldsymbol{R}$ over the angle $\theta$.

Finally, we take the scalar order parameter $S$ as the maximum eigenvalue of $<\boldsymbol{Q}_p>$ and its eigenvector $\hat{\boldsymbol{e}}$ as the averaged cell director.

Similarly, we define the bipolar order parameter $S_b$ as following. First, we use the coordinates of boundary cells to reconstruct the biofilm-gel interface. Next, for every single outmost cell, we define the local surface normal $\boldsymbol{n}_{\text{norm}}$. Also, we define the position vector $\boldsymbol{r}_i$ by calculating the position of each cell $i$ relative to the biofilm center. The bipolar order parameter $S_b$ is defined as $S_b = 1/2(3|\boldsymbol{n}_i' \cdot \boldsymbol{r}_i'| - 1)$, where $\boldsymbol{n}_i'$ and $\boldsymbol{r}_i'$ are the normalized projection vector of $\boldsymbol{n}_i$ and $\boldsymbol{r}_i$ onto the local tangent plane defined by $\boldsymbol{n}_{\text{norm}}$, respectively. $S_b$ is averaged over three outmost layers of cells to reduce randomness.

Biofilm growth dynamics and morphology from ABM

As shown in **Fig. S2**, our model is able to reproduce the growth dynamics of G-II biofilm from the early stage to the mature state: Starting with a single cell lying on the glass surface, cells first proliferate and form an 2D layer. When the local in-plane pressure accumulates and reaches the threshold for verticalization instability (*4*), cells tend to be vertical then release part of the growth pressure, and this process generates the initial out-of-plane growth for the transition from 2D expansion to 3D growth of a biofilm. Biofilms deform the surrounding gel during its expansion, resulting different level of growth-induced stress which depends on the biofilm volume and gel stiffness. The stiffness-dependent morphology change is also quantitively captured by our agent-based model. We tune the Young's modulus of surrounding gel from $10^1$ to $10^4$ Pa and find a sharp transition in biofilm shape around $E_{\text{gel}} = 10^2$ Pa, quantitively reproducing the experimentally observed domes-to-lenses shape transition. More detailed phase diagram of contact angle about Young's modulus of gel and the biofilm-substrate friction can be also found in the reference (*12*).

G-II biofilm simulations

We have shown the spatiotemporal evolution of hydrostatic pressure, equivalent shear stress, density and rotational speed in the G-I biofilms. Here for completeness and further verification of our hypothesis, the same physical quantities are visualized in **Fig. S5** and **Fig. S6**. As shown in **Fig. S5**, the spatial distribution of stresses shares similar characteristics with the G-I biofilm. Specifically, the pressure and shear stress follow the same trends from the inner region of G-II biofilm to the outer region. Near the center, the pressure is the highest and the equivalent shear stress is relatively low, and near the biofilm-gel interface, the pressure goes down, but the equivalent shear stress reaches its maximum. Since G-II biofilm has same experimental settings except the existence of the rigid glass substrate, the similarity between the pressure and shear stress distribution of G-I and G-II biofilms can be explained by regarding the glass substrate as a plane of symmetry to the first order. However, the existence of the rigid substrate slightly changes the first principal stress direction of the bottom layer of G-II biofilm, from randomly oriented in x-y plane to mostly in z direction, due to cell verticalization (*13-15*).

Effects of cell stiffness

As a computational exploration for investigating the effects of cell behaviors to the spatiotemporal evolution of cell ordering, we alter the soft-shell cell stiffness $E_0$ from relatively soft (~100 Pa) to relatively rigid (~5000 Pa) and keep other simulation setting unchanged. Previous study (*16*) has shown the cell stiffness is related to the average size of local aligned group. For the G-I biofilm, we define the bipolarly aligned boundary layer using the following method: Based on the reconstructed biofilm-gel interface, we define a series of self-similar ellipsoid surfaces by the interval of long-axis $\Delta L = 0.2 \mu m$, as the possible inner surface of the boundary layer. Then we increase the thickness by the increment $\Delta L$ and calculate the averaged bipolar order parameter $S_b$ for all cells between the biofilm-gel interface and the given inner surface. The region where $S_b > 0.4$ is regarded as the bipolar boundary layer, and its thickness is defined as the length difference of two long axis of the boundaries. As shown in **Fig. S7**, we indeed observe a two-fold change in the thickness of boundary aligned layer when changing the $E_0$ from 100 Pa to 5 kPa, which can be explained as the increase of energy cost for neighbor cells to have nonparallel configuration and overlap.

Numerical experiments of lateral pressure

As shown in **Fig. 5**, to further illustrate the bidirectional coupling effects between cell ordering and stresses, we design a numerical experiment by imposing an artificial compression on lateral direction. The G-II biofilm is first growing under normal condition (without lateral pressure) for ~11 hrs; then we impose lateral pressure by biaxially deforming the surrounding gel boundary by $\frac{\Delta L_x}{L_x} = 0.2$ and $\frac{\Delta L_y}{L_y} = 0.2$. Denote the time point imposing lateral compression as $t_0$, we measure the evolution of average shear stress $\tau_{eq}$ and the alignment parameter $\alpha = |\hat{n}_1 \cdot \hat{n}_c|$ during the time window $-0.2 < \frac{t-t_0}{T_{\text{double}}} < 1$. We find that the lateral compression nearly instantly changes the stress state across the whole biofilm. In contrast, the reorientation process takes roughly $\frac{T_{\text{double}}}{2}$ to reach a steady value, indicating the existence of local energy barriers for each cell due to the configuration of neighboring particles.

Numerical experiments with spatially patterned gel stiffness

We design a variation of G-II biofilm simulation by spatially patterning the surrounding gel stiffness. We set the gel modulus 10-fold softer inside a cylindrical region, with the center line pass through the initial seeding cell and the radius of 20 μm. The remaining part of gel has the homogeneous Young's modulus of 20 kPa. Under this confinement settings, we observe significant morphology change and different cell alignment compared with normal simulation of G-II biofilms. Namely, the part of biofilm under the soft region forms a "bleb" indicating non-uniform indentation depth of the soft part of the gel. Also, compared with normal G-II biofilms, significant number of cells are verticalized due to altered first principal stress direction. Our numerical experiments demonstrate the possibility of mechanically controlling biofilm morphology and cell ordering, and it might lead to more precise spatiotemporal control of the stress field and cell orientation field inside the biofilms.

Details of experimental measurements

*Bacterial strains and cell culture*

The details of bacterial strains can be found in the reference (*12*). The biofilm growth experiments begin by first growing *V. cholerae* cells in LB broth (BD) overnight under shaken conditions, then back-diluted 30× in M9 media and grown under shaken conditions until reaching an optical density (OD) of 0.05-0.25. Different concentrations of agarose polymer are boiled in M9 media and then placed in a water bath to cool to 40-50°C without gelation. A 1 μL droplet of the bacterial culture is placed in the center of a glass-bottomed 96 well plate (MatTek) after being diluted in M9 media to an OD of 0.001-0.003. The bacteria are sandwiched between the solidified gel and the glass substrate by the 20 μL of liquid agarose that is used to cover the droplet. (Note that we ignore the droplet's ~5% dilution of the agarose.) In order to serve as a nutrient reservoir, 200 μL of M9 media is put in the well on top of the solidified agarose. Finally, cells are cultivated in static conditions at 30°C and imaged throughout several developmental stages.

*Overview of image analysis*

Raw images are first deconvolved using Huygens 20.04 (SVI) using a measured point spread function. The deconvolved three-dimensional confocal images are then binarized, layer by layer, with a locally adaptive Otsu method. To accurately segment individual bacterium in the densely packed biofilm, we develop an adaptive thresholding algorithm. Once segmented, we extract the cell positions by finding the center of mass of each object, and the cell orientations by performing a principal component analysis. The positions and directions of each cell are converted from cartesian $(x, y, z, \hat{n}_x, \hat{n}_y, \hat{n}_z)$ to cylindrical polar $(r, \psi, z, \hat{n}_r, \hat{n}_\psi, \hat{n}_z)$ coordinates where the origin is found by taking the center of mass of all of the segmented cells in the $(x, y)$ plane. Reconstructed biofilm images are rendered using Paraview.

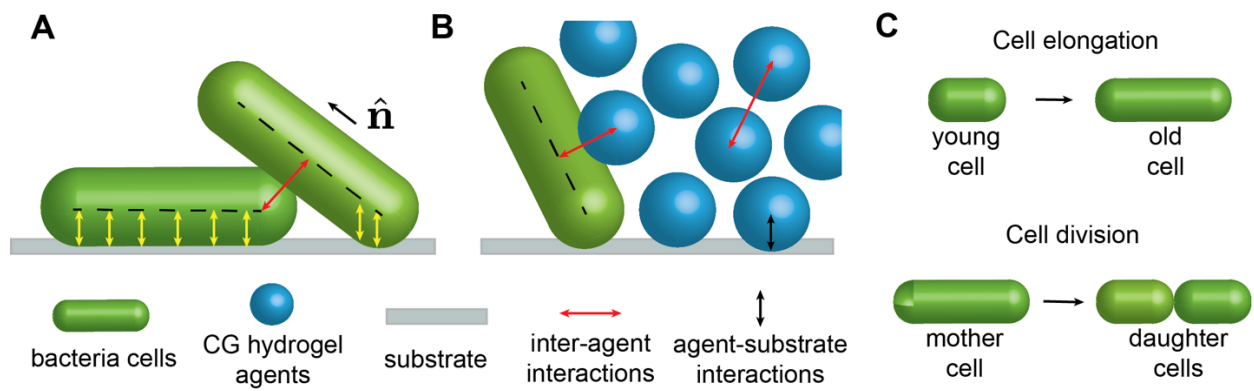

**Fig. S1. A schematic illustration of agent-based model. A.** Cell-cell interactions. The unit vector $\hat{n}$ represents the director of a single cell. **B.** Cell-gel interactions. **C.** Schematics of cell growth and division. After cell division, the mother cell is replaced by two daughter cells with nearly equal length.

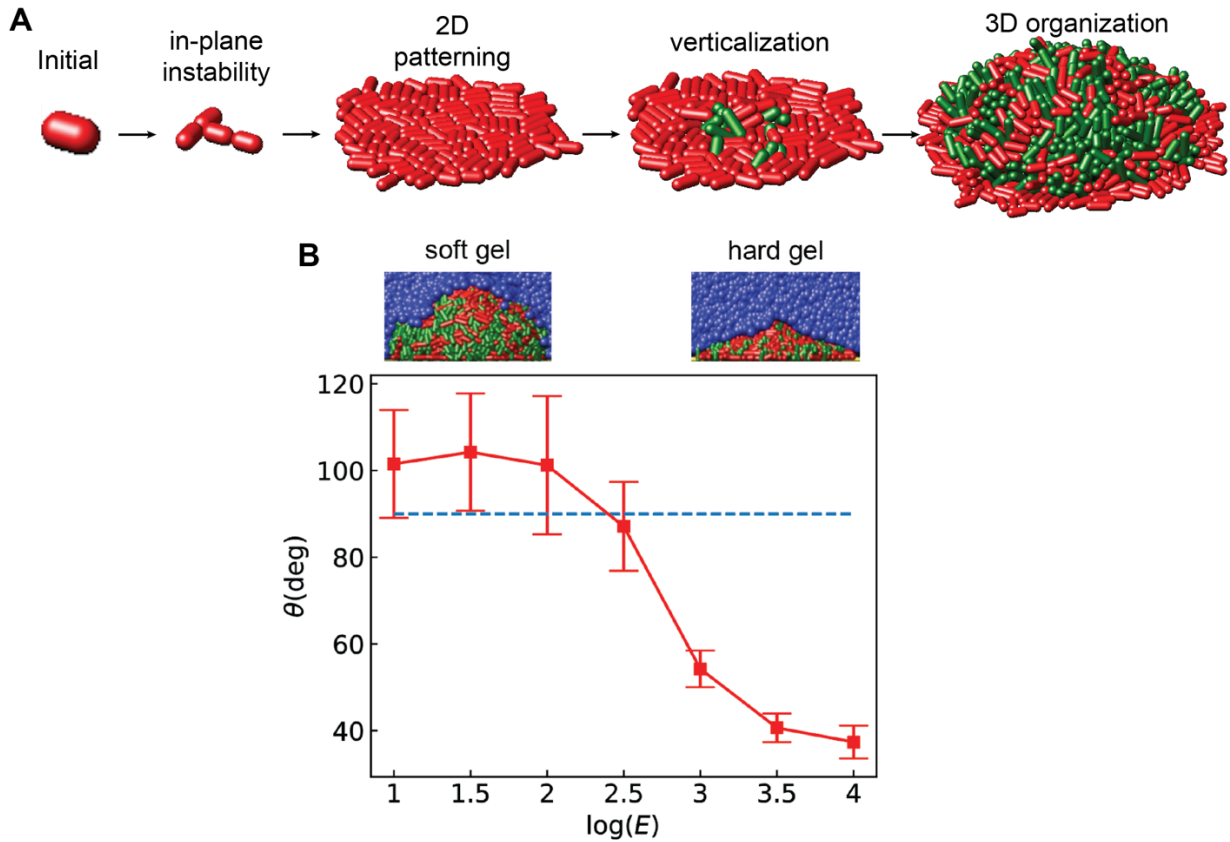

**Fig. S2. Agent-based model captures biofilm growth morphodynamics. A.** Representative G-II biofilm formation process given by agent-based simulations. **B.** Contact angle changes with the Young's modulus of the surrounding gel. Blue dashed line: 90 degrees. Top subfigures: the cross-section view of the mature (grown after 12 hrs.) G-II biofilm in soft gel ($E_{gel} \sim 10^2$ Pa) and hard gel ($E_{gel} \sim 10^4$ Pa).

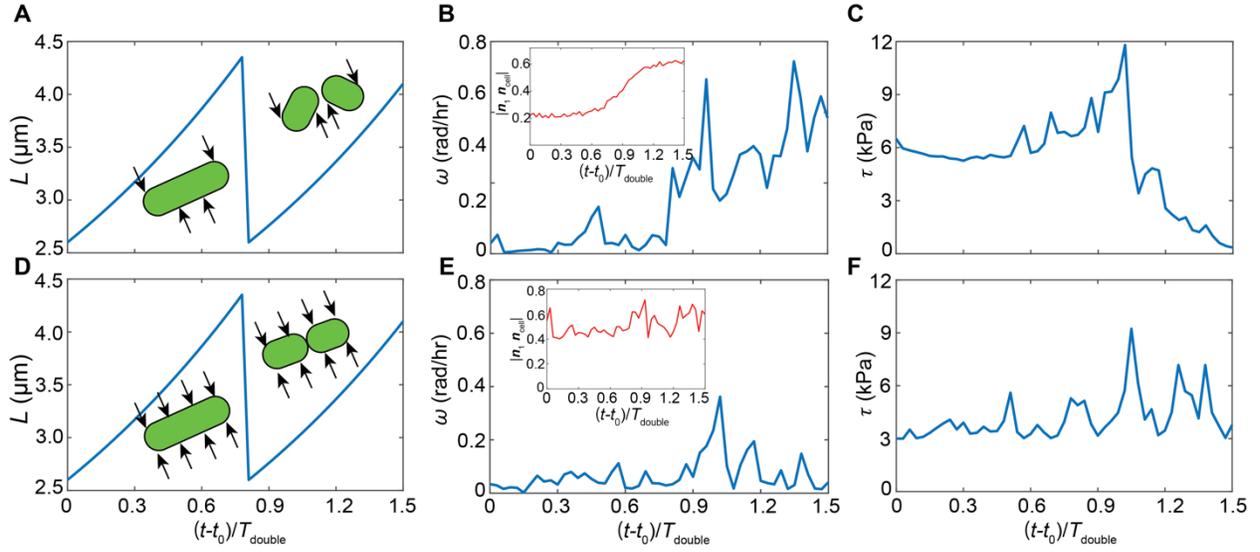

**Fig. S3. Representative traces for unstable (A-C) and stable (D-F) configurations of a cell after division. A.** Time evolution of single cell length before/after cell division. Inset: Schematic illustration of mechanical instability after cell division. **B.** Time evolution of the cell rotation speed before/after cell division. Inset: Time evolution of alignment $|\hat{n}_1 \cdot \hat{n}_c|$. **C.** Time evolution of the equivalent shear stress. **D-F.** Representative traces of cell length (**D**), rotation speed (**E**) and shear stress (**F**) for a stable configuration after division, respectively.

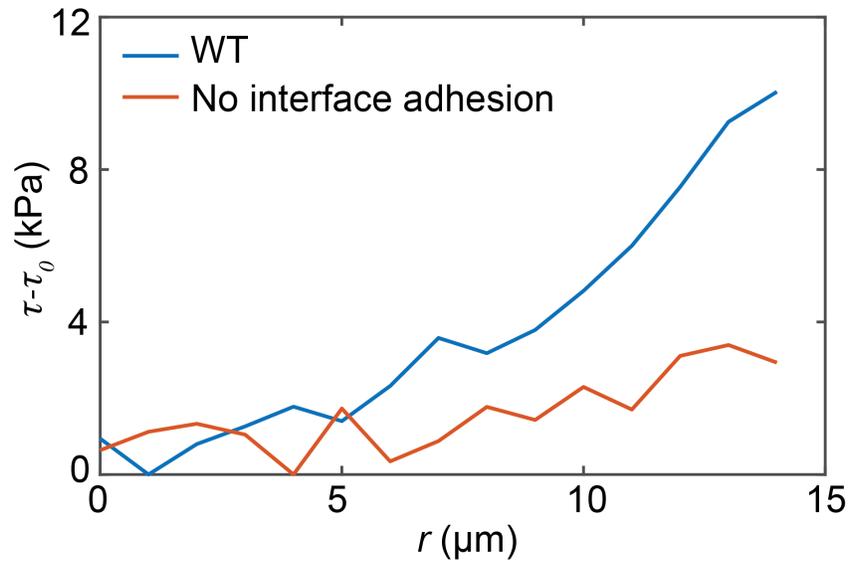

**Fig. S4. Transmission of boundary shear stress.** Compared with WT biofilm (with normal interface adhesion), biofilms without interface adhesion have significantly less boundary shear stress. The distance $r$ is defined by the long axis length of self-similar ellipsoids.

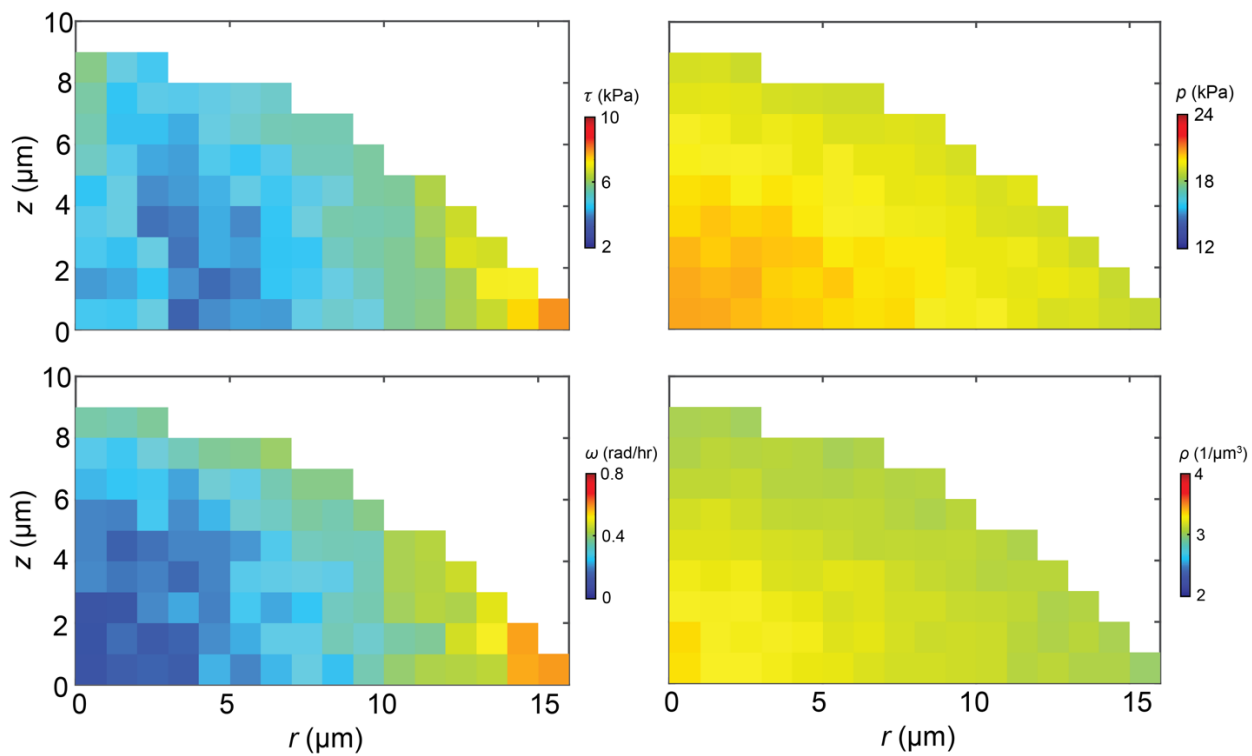

**Fig. S5. Spatial distribution of stresses, density and rotation speed of G-II biofilms. A.** Spatial distribution of the equivalent shear stress. **B.** Spatial distribution of hydrostatic pressure. **C.** Rotational speed. **D.** Cell density.

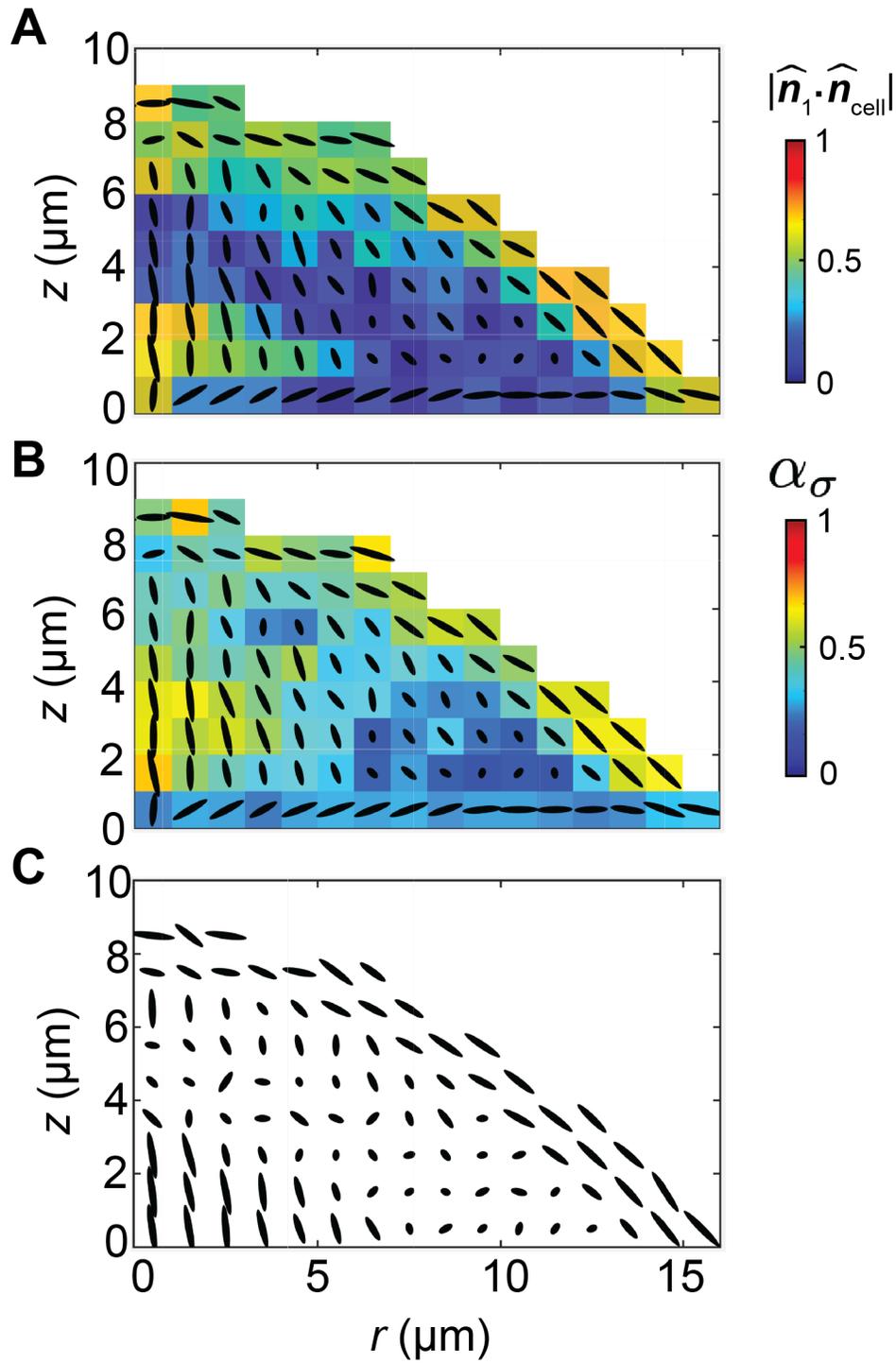

**Fig. S6. Spatial distribution of alignment $|\hat{\boldsymbol{n}}_1 \cdot \hat{\boldsymbol{n}}_c|$, stress anisotropy and direction of first principal stress of G-II biofilms. A.** Alignment $|\hat{\boldsymbol{n}}_1 \cdot \hat{\boldsymbol{n}}_c|$. **B.** Stress anisotropy $\alpha_\sigma$. **C.** Direction of first principal stress.

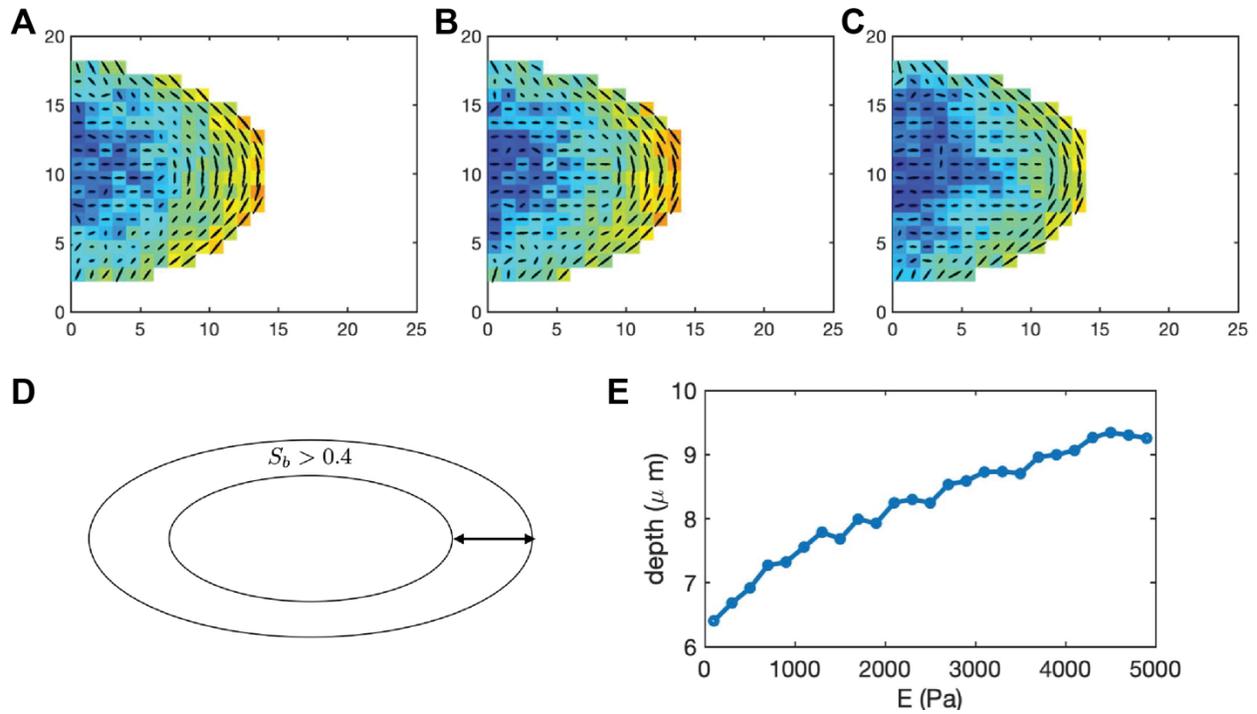

**Fig. S7. The effect of cell stiffness on the thickness of boundary layer. A-C.** Three representative simulations for $E_0 = 150$ Pa **(A)**, $E_0 = 1200$ Pa **(B)** and $E_0 = 4000$ Pa **(C)** cell stiffnesses. Black oval denotes cell orientations and color denotes the aligning order parameter $S$. **D.** Schematics showing the definition of the boundary bipolar layer. **E.** The relation between the depth of boundary bipolar layer and the cell stiffness.

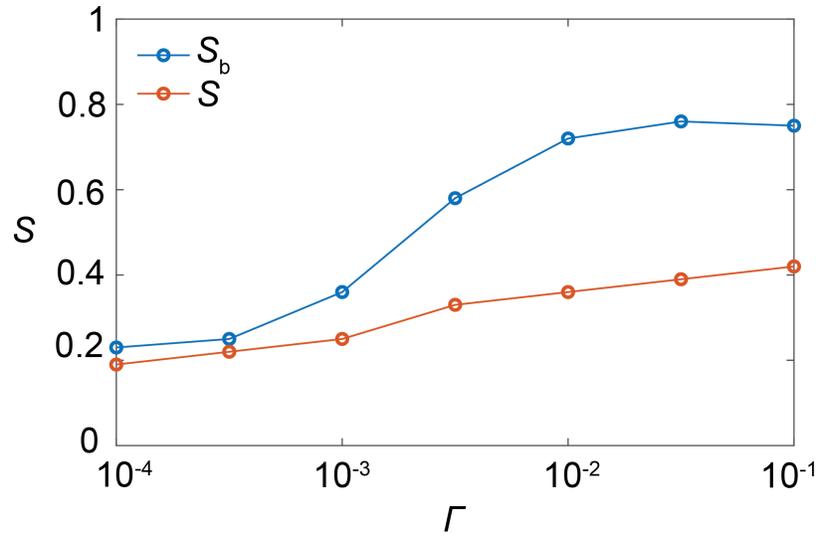

**Fig. S8. Biofilm-gel interfacial adhesion controls boundary cell ordering.** Blue solid line: The boundary bipolar ordering $S_b$ changes with the interfacial adhesion $\Gamma$. Orange solid line: the overall ordering $S$ also has a slight increase with $\Gamma$, mainly due to the increase of the boundary ordering.

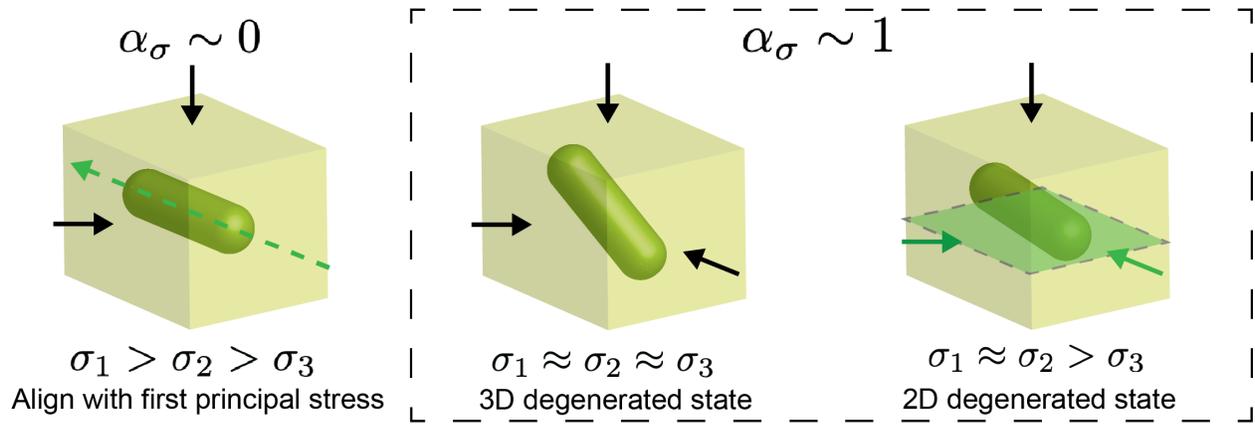

**Fig. S9. Schematics illustration of possible stress states in biofilms.** Black and green arrows denote the directions of the principal stresses. Green plane with black dashed boundary represents the degenerate plane of minimal compression.